# Dual-Polarization OFDM-OQAM Wireless Communication System

Hosseinali Jamal, David W. Matolak
Department of Electrical Engineering
University of South Carolina
Columbia, SC, USA
hjamal@email.sc.edu, matolak@cec.sc.edu

*Abstract*—In this paper we describe the overall idea and results of a recently proposed radio access technique based on filter bank multicarrier (FBMC) communication system using two orthogonal polarizations: dual-polarization FBMC (DP-FBMC). Using this system we can alleviate the intrinsic interference problem in FBMC systems. This enables use of all the multicarrier techniques used in cyclic-prefix orthogonal frequency-division multiplexing (CP-OFDM) systems for channel equalization, multiple-input/multiple-output (MIMO) processing, etc., without using the extra processing required for conventional FBMC. DP-FBMC also provides other interesting advantages over CP-OFDM and FBMC such as more robustness in multipath fading channels, and more robustness to receiver carrier frequency offset (CFO) and timing offset (TO). For DP-FBMC we propose three different structures based on different multiplexing techniques in time, frequency, and polarization. We will show that one of these structures has exactly the same system complexity and equipment as conventional FBMC. In our simulation results DP-FBMC has better bit error ratio (BER) performance in dispersive channels. Based on these results, DP-FBMC has potential as a promising candidate for future wireless communication systems.

## I. Introduction

OFDM with cyclic prefix (CP) is the most widespread multicarrier communication system due to its simplicity and robustness against frequency selective channels. But it loses spectral efficiency in highly-dispersive channels where a large portion of each OFDM symbol should be added as CP. Also, the OFDM power spectrum is not compact due to the *sin(x)/x (sinc)* frequency response of its rectangular pulses, therefore it requires windowing techniques and a large number of guard subcarriers to reduce the out-of-band (OOB) power emission. As an alternative, FBMC has been proposed to increase the spectral efficiency via a more compact power spectral density (PSD) [1], [2]. FBMC's non-rectangular pulse shapes do not require a CP and therefore FBMC can achieve larger spectrum efficiency, and lower OOB power.

In this paper we investigate the widespread FBMC technique based on Saltzberg's method [3]. In recent years this system has seen much interest. This communication system is also known as staggered multitone (SMT) FBMC or OFDM-offset quadrature amplitude modulation (OQAM) [1]. This technique enables symbol-rate spacing between adjacent subcarriers without inter-symbol interference (ISI) and inter-carrier interference (ICI) in distortionless channels. This works by a shift of half the symbol period between in-phase and quadrature components of QAM symbols [3]. In FBMC, because of the real nature of the symbols, the orthogonality condition is only satisfied in the real domain, and in dispersive channels where the channel response is in general complex, FBMC suffers from imaginary *intrinsic interference*. In order to mitigate this interference we require an interference cancellation method. In the literature we find several methods for mitigating this interference, but almost all of these techniques add complexity [4]-[6].

Using polarization-division multiplexing (PDM) we can multiplex signals on electromagnetic waves of two orthogonal polarizations on the same carrier frequency without any interference, assuming perfect (infinite) cross-polarization discrimination (XPD). The XPD is the ratio of the signal amplitudes on the desired polarization to that on the opposite polarization. In [7] this technique has been reported in satellite television links to double the throughput. It has also been studied for fiber optic communication links using two orthogonal left- and right-hand circularly polarized light beams in the same light guide fiber at the same frequency [8]. Using PDM in terrestrial and vehicular wireless communication environments to double the capacity is nearly impossible when platforms are mobile, due to the non-stability of antenna position and often rich scattering in the wireless channels.

In this paper we use the PDM technique and propose dual-polarization FBMC (DP-FBMC) not to double the capacity but instead to solve the intrinsic imaginary interference problem of FBMC systems: by using two polarizations we add an extra dimension to suppress the intrinsic interference. We will show that DP-FBMC has the advantage of significantly suppressing this interference thanks to the polarization multiplexing and the use of proper prototype filters.

This paper is organized as follows: in Section II we describe the conventional OFDM-OQAM system model. In Section III we describe the proposed DP-FBMC communication system, and in Section IV we provide simulation results for bit error ratio (BER) performance comparing CP-OFDM, FBMC, and DP-FBMC in ITU pedestrian A and vehicular A channel models. We also compare the peak-to-average power ratio (PAPR) and PSD of these systems, and evaluate DP-FBMC performance versus carrier frequency and timing offsets, as well as BER performance in low XPD conditions. In Section V we provide conclusions and briefly describe future work.

## II. CONVENTIONAL OFDM-OQAM

In OFDM-OQAM, real valued OQAM symbols at subcarrier index $n$ and symbol index $m$, $a_{n,m}$ are filtered and modulated on different subcarriers, as described by the following equation,

$$s(t) = \sum_{n=0}^{M-1}\sum_{m\epsilon\mathbb{Z}} a_{n,m} h\left(t - m\frac{T_0}{2}\right) e^{\frac{j2\pi nt}{T_0}} e^{j\theta_{n,m}}. \quad (1)$$

where $h(t)$ is the prototype filter with length $KT$, and $K$ is termed the overlapping factor. The exponential part in this equation modulates the filtered OQAM symbols onto $M$ subcarriers with $1/T_0$ frequency spacing, along with an extra phase component $\theta_{m,n} = e^{j\frac{\pi}{2}(m+n)}$ to add the $\pi/2$ phase shifts to adjacent symbols in order to satisfy the real orthogonality condition [2]. Based on this structure, OFDM-OQAM symbols are offset or overlapped by $T_0/2$ in time to yield maximum spectral efficiency. Relative to CP-OFDM, one main shortcoming of OFDM-OQAM is its complexity. In order to reduce this complexity we can use polyphase network (PPN) structures at both transmitter and receiver and use fast and inverse fast Fourier transforms [2] as shown in Figure 1.

For details regarding PPN structures and IFFT/FFT based OFDM-OQAM implementations refer to [2], [10]. In the FBMC transmitter, after the IFFT block, subcarriers will be filtered through the PPN network. For each block of $M$ input subcarriers, the parallel to serial (P/S) output sequence is a signal vector with the same length as the prototype filter. These signal sequences are then overlapped at $T_0/2$ intervals. The reverse process is followed at the receiver. In order to illustrate the transmitting OQAM symbols in time and frequency we use the time-frequency phase-offset diagram. In Figure 2 we show this diagram for OFDM-OQAM. In this structure the solid circles and squares denote the $\pi/2$ phase shifts between adjacent symbols (adjacent in both time and frequency domains). In the following equations we analyze the intrinsic imaginary interference in FBMC.

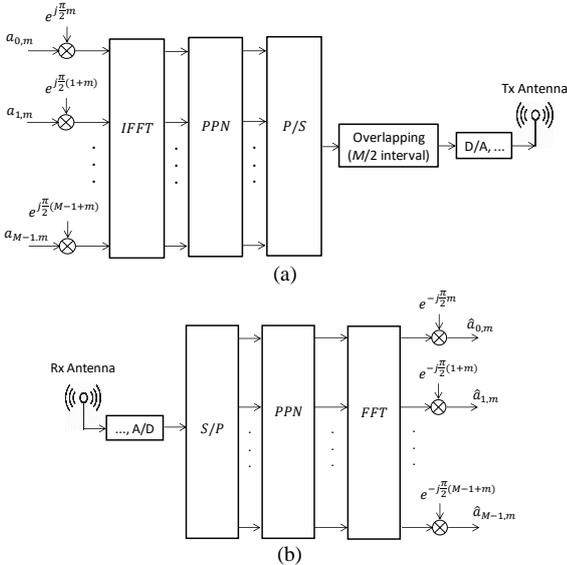

Figure 1. OQAM-OFDM communication system; (a) transmitter, (b) receiver.

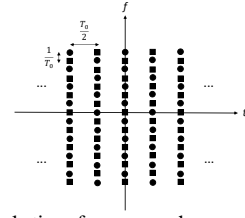

Figure 2. FBMC symbols time-frequency phase-space lattice ($M$=16).

First we re-write (1) as follows,

$$s(t) = \sum_{n=0}^{M-1}\sum_{m=-\infty}^{\infty} a_{n,m} v_{n,m}(t) \quad (2)$$

where,

$$v_{n,m}(t) = h\left(t - m\frac{T_0}{2}\right) e^{j\frac{2\pi nt}{T_0}} e^{j\theta_{n,m}} \quad (3)$$

In (3), $v_{n,m}(t)$ represents the time shifted and frequency modulated versions of the prototype filter response $h(t)$. Here the real orthogonality condition of OFDM-OQAM system can be expressed as,

$$\Re\{\langle v_{n,m}, v_{p,q}\rangle\} = \Re\left\{\int_{kT_0}^{(k+1)T_0} v_{n,m}(t)v_{p,q}^*(t)dt\right\} = \delta_{n,p}\delta_{m,q} \quad (4)$$

where $\delta_{n,p}$ is the Kronecker delta function, equal to one if $n=p$ and 0 if $n\neq p$, and $\Re$ is the real part of the quantity inside the braces, and $k$ is any integer. In dispersive multipath channels we can write the received symbols as follows,

$$\hat{a}_{n,m} = H_{n,m}(a_{n,m} + jI_{n,m}) + Z_{n,m} \quad (5)$$

where $H_{n,m}$ are the complex channel transfer function samples at subcarrier $n$ and symbol $m$, $I_{n,m}$ is the so called intrinsic interference, and $Z_{n,m}$ is the additive white Gaussian noise (AWGN) variable. As long as the $H_{n,m}$ values are complex, the real valued transmitted symbols will suffer from the presence of the intrinsic interference $I_{n,m}$ at the receiver. Therefore we need to find a way to remove the $I_{n,m}$. In [4] the authors proposed an interesting technique to remove this interference based on "auxiliary pilots." In this paper we also use this technique for channel equalization in the conventional FBMC system. To calculate $I_{n,m}$ we first define the following prototype filter time-frequency localization function,

$$v_{n,m}^{p,q} = -j\langle v_{n,m}, v_{p,q}\rangle = -j\left\{\int_{kT_0}^{(k+1)T_0} v_{n,m}(t)v_{p,q}^*(t)dt\right\} \quad (6)$$

From (4) and (5) we note that (6) are purely real values for $n\neq p$ or $m\neq q$. Now we can calculate the $I_{n,m}$ knowing the adjacent real OQAM symbols from,

$$I_{n,m} = \sum_{(p,q)\in A_{\Delta n,\Delta m}} a_{n+p,m+q} v_{n,m}^{p,q} \quad (7)$$

where,

$$A_{\Delta n,\Delta m} = \{(p,q), |p|\leq \Delta n, |q|\leq \Delta m\backslash\{0,0\} \mid H_{n+p,m+q} \cong H_{n,m}\} \quad (8)$$

Now by defining a reference symbol $a_{0,0}$, $A_{\Delta n,\Delta m}$ is the set of nearby indices with frequency within $\Delta n$ subcarriers and time within $\Delta m$ symbols of the reference subcarrier and symbol indices (*n*=0, *m*=0). In practice, using well-localized prototype filters, $\Delta n, \Delta m$ can be selected as small as one [5]. This means that most of the intrinsic interference results from the adjacent symbols and interference from symbols outside the $A_{\Delta n,\Delta m}$ neighborhood is negligible. Of course in our analysis as we describe in detail in [13], equations (5)-(8) rely on the assumption of channels with low frequency selectivity, relying in a flat fading model for the subchannels.

III. PROPOSED DP-FBMC SYSTEM

In this section we describe the proposed DP-FBMC communication system from [13]. In this DP-FBMC proposal we define three different DP-FBMC multiplexing structures. In Figure 4 diagrams we show the time-frequency-polarization phase-lattice structure of all DP-FBMC structures. *Structure I* is Based on time-polarization division multiplexing (TPDM). In this method we isolate adjacent symbols on two orthogonal polarizations by time multiplexing of adjacent OQAM symbols. This approach removes the intrinsic interference that results from adjacent symbols (assuming perfect cross-polarization), but some of this interference still exists on adjacent subcarriers during each symbol period.

In *Structure II* we use a frequency-division multiplexing technique (FPDM) as shown in Figure 3(b). In this approach we isolate the adjacent transmitted subcarriers on the two polarizations by frequency multiplexing of symbols. Figure 3(c) shows the DP-FBMC *Structure III* based on time-frequency-polarization division multiplexing (TFPDM). In this structure using both time and frequency multiplexing we separate half the subcarriers on both polarizations at every symbol period, and the order of subcarriers changes at subsequent symbol periods.

To better understand the DP-FBMC idea, in Tables 1 to 3 we provide numerical examples of the intrinsic interference $v_{n,m}^{p,q}$ values surrounding every reference symbol for the filters based on isotropic orthogonal transform algorithm (IOTA) technique [15], PHYDYAS [9], and square-root raised cosine (SRRC) with overlapping factor $K = 4$, respectively. In these tables we colored the symbol locations based on DP-FBMC *Structure III*. Here the red samples represent the time-frequency filter response on the same polarization as the subject symbol, and the black symbols are on the other polarization with no interference to the subject symbol (again assuming perfect polarization isolation). As we notice from this table the interference caused from adjacent subcarriers is suppressed entirely but there are still symbols (red) on the same polarization that can cause interference. For the SRRC filter, via some numerical trials, we determined heuristically that a roll-off factor $\alpha = 2/K$ is a good choice and we chose alpha factor one for IOTA. As can be seen, from Tables 1 to 3, the red intrinsic interference sample values decreased, and as will be shown, SRRC filters will yield better BER performance. To obtain even lower intrinsic interference values with the SRRC, we can increase the overlapping factor at the cost of slightly increasing complexity and frame length. In Figure 4 we plot the normalized impulse responses of these filters.

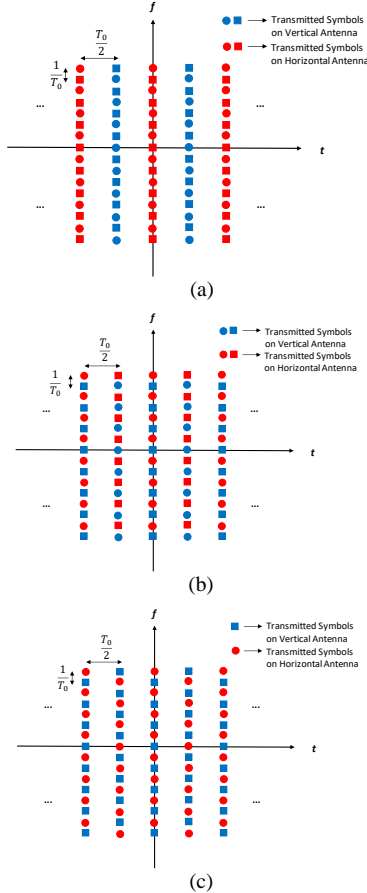

Figure 3. DP-FBMC structures, (a) *Structure I* based on TPDM, (b) *Structure II* based on FPDM, (c) *Structure III* based on TFPDM.

Table 1. IOTA $v_{0,0}^{p,q}$ values for *p* = [-2, 2], *q* = [-3, 3] and *K* =4.

| p\q | -3 | -2 | -1 | 0 | 1 | 2 | 3 |
|---|---|---|---|---|---|---|---|
| -2 | -0.0194j | 0 | 0.0413j | 0 | -0.0413j | 0 | -0.0194j |
| -1 | 0.0116j | 0.0413j | 0.2327j | 0.4378j | 0.2327j | 0.0413j | 0.0116j |
| 0 | 0.0194j | 0 | 0.4380j | $a_{0,0}^{0,0}$=1 | -0.4380j | 0 | -0.0194j |
| 1 | 0.0116j | -0.0413j | 0.2327j | -0.4378j | 0.2327j | -0.0413j | 0.0116j |
| 2 | 0 | 0 | 0.0413j | 0 | -0.0413j | 0 | 0 |

Table 2. PHYDYAS $v_{0,0}^{p,q}$ values for *p* = [-2, 2], *q* = [-3, 3] and *K* = 4.

| p\q | -3 | -2 | -1 | 0 | 1 | 2 | 3 |
|---|---|---|---|---|---|---|---|
| -2 | -0.0644j | 0 | 0 | 0 | 0 | 0 | -0.0644j |
| -1 | 0.0442j | 0.1250j | 0.2058j | 0.2393j | 0.2058j | 0.1250j | 0.0442j |
| 0 | -0.0644j | 0 | 0.5645j | $v_{0,0}^{0,0}$=1 | -0.5645j | 0 | -0.0644j |
| 1 | 0.0442j | -0.1250j | 0.2058j | -0.2393j | 0.2058j | -0.1250j | 0.0442j |
| 2 | 0 | 0 | 0 | 0 | 0 | 0 | 0 |

Table 3. SRRC $v_{0,0}^{p,q}$ values for *p* = [-2, 2], *q* = [-3, 3] and *K* =4.

| p\q | -3 | -2 | -1 | 0 | 1 | 2 | 3 |
|---|---|---|---|---|---|---|---|
| -2 | -0.1122j | 0 | 0 | 0 | 0 | 0 | -0.1122j |
| -1 | 0.095j | 0.1263j | 0.15j | 0.1589j | 0.15j | 0.1260j | 0.095j |
| 0 | 0.1122j | 0 | 0.6015j | $v_{0,0}^{0,0}$ | -0.6015j | 0 | -0.1122j |
| 1 | 0.095j | -0.1263j | 0.15j | -0.1589j | 0.15j | -0.1260j | 0.095j |
| 2 | 0 | 0 | 0 | 0 | 0 | 0 | 0 |

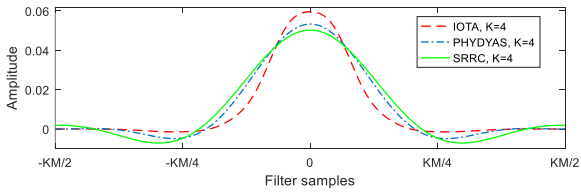

Figure 4. IOTA, PHYDYAS, and SRRC prototype filters impulse responses with $K = 4$.

In [13] we provide analysis of complexity for all DP-FBMC structures. As a quick complexity review first we note that for DP-FBMC *Structures II* and *III* we need a second IFFT and FFT at both transmitter and receiver, therefore they have higher complexity than conventional FBMC. For *Structure I*, as long as we can share the same IFFT/FFT at every symbol period and polarization we obtain the same complexity and equipment as conventional FBMC. In this paper we focus on *Structure I* in simulations and analysis; more results and analysis are provided in [13]. For DP-FBMC systems, as long as the channel equalization process is being done separately on both polarizations, the equalization complexity is higher than for conventional FBMC, and as will be shown, this is the cost for better performance.

## IV. SIMULATION RESULTS

We compare the BER performance of CP-OFDM, conventional FBMC, and DP-FBMC via simulations. Note that in our simulation first we provide results assuming perfect XPD, and at the end results for some realistic XPD values.

In our simulations following the approach in [13] we evaluate BER performance in two multipath fading channels. We compare PAPR, and the effects of carrier time and frequency offsets. We also show the PSD of DP-FBMC using different prototype filters and overlapping factors. These simulations were performed at baseband with no channel coding. We have $M = 512$ subcarriers with 16 symbols on each frame, and the signal bandwidth $B = 10$ MHz. hence the temporal symbol interval is $T_0 = 512/10^7 = 51.2$ µs. As mentioned in [13], the mean RMS-DS values for pedestrian and vehicular A channels are 46 and 370 ns, respectively. For channel models we chose pedestrian and vehicular channels A from ITU-R recommendation M.1225 [10]. In these simulations we assume Ricean fading with Rice factor 10 dB for the pedestrian and Rayleigh fading for the vehicular. For CP-OFDM we chose $M/16$ of a symbol period for the CP for both pedestrian channel A, and vehicular channel A. For conventional FBMC we chose the PHYDYAS prototype filter with $K = 4$. We use 33 subcarriers for guard band (17 on the left and 16 on the right of the signal spectrum), and also null the DC subcarrier. For channel estimation we use scattered-pilot symbols on 30 equally spaced subcarriers with 4 symbols temporal spacing. For channel equalization, we used the least squares (LS) and discrete Fourier transform (DFT)-based channel estimation noise reduction and spline interpolation technique [12]. In FBMC we used the auxiliary pilot technique based on [4]. For all communication systems we assume perfect timing offset estimation and synchronization.

In Figures 5(a) and 5(b) we plot BER vs. $E_b/N_0$ for 16-QAM and QPSK modulations in the pedestrian A and vehicular A channels, respectively. As explained in [13], *Structures I* and *III* have similar BER results, therefore we only show the results using *Structure I*. According to these results DP-FBMC has better BER performance than CP-OFDM and conventional FBMC after channel equalization for highly dispersive channels (Vehicular channel). Here we also plot BER results with perfect channel knowledge and zero-forcing technique.

This better DP-FBMC BER is due to larger temporal spacing of each of the polarization's symbols. In DP-FBMC *Structures I* and *III* because of the time multiplexing, on both structures the symbol spacing increases compared to the symbol spacing in conventional FBMC. This yields lower ISI in multipath channels and hence lower BER. Also, as can be seen using the IOTA filter slightly worsens the BER. This is because using DP-FBMC *Structure I*, the interferences caused from adjacent symbols and adjacent subcarriers are equal if the IOTA pulse shaping filters are used. Here we note that IOTA filter has almost same result (results are not shown here) compared to other filters in DP-FBMC *Structure III*, because in *Structure III* the symbols are multiplexed both in time and frequency domains. We also note that as expected for higher overlapping factors, SRRC filter results better BER results.

Figure 6(a) provides PAPR complementary cumulative distribution function results for these systems. For FBMC systems we assume frame structures without end tails caused from filter tails from the overlapping factor. In [14] we study the PAPR of DP-FBMC systems in detail. As can be seen, the DP-FBMC *Structure I* has PAPR very close to that of conventional FBMC and CP-OFDM for larger overlapping factors such as $K = 8$. Here we also note that DP-FBMC Structures *II* and *III* have exactly the same PAPR as CP-OFDM and conventional FBMC with different overlapping factors, because temporally all the symbols are present the entire frame period.

PSDs appear in Figure 6(b). Note that in these results similar to PAPR, we truncate two ends of the frame waveforms due to filter tails in order to make the frame lengths as short as possible. We can observe that when DP-FBMC has the same $K$ as conventional FBMC, the PSD is nearly identical. As expected, with SRRC as $K$ increases, the OOB power decreases.

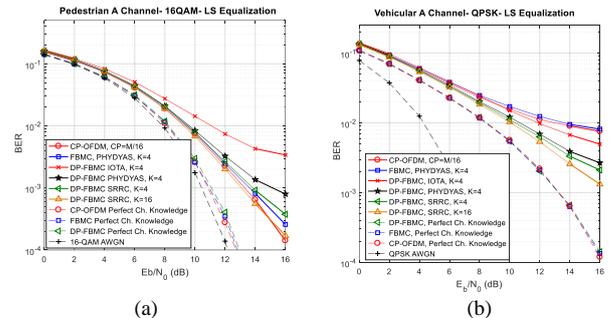

Figure 5. BER vs. $E_b/N_0$, (a) 16-QAM, ITU-R pedestrian A channel, (b) QPSK, ITU-R vehicular A channel

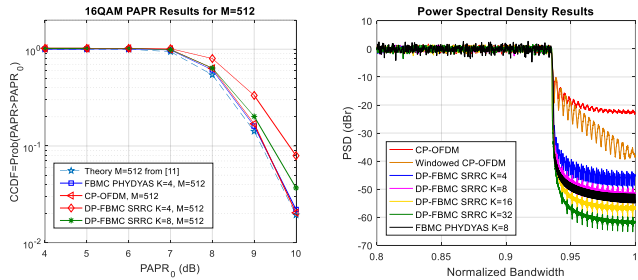

Figure 6. (a) *CCDF* vs. *PAPR₀*, (b) PSD vs. normalized bandwidth

Figures 7(a) and 7(b) plot simulated BER versus carrier frequency (*CFO*) and timing offset ($n_\tau$) at the receiver, respectively, in an AWGN channel, with 16-QAM modulation, SNR = 12 dB and *M* = 512 subcarriers. The CFO values are normalized to the subcarrier bandwidth and $n_\tau$ is the number of samples offset. Channel bandwidth *B*=10 MHz. These results confirm that DP-FBMC has better performance in the presence of time and frequency offsets (and again, larger overlapping factors slightly enhance the performance).

In Figure 8 assuming orthogonal circular polarizations, we plot the 16-QAM BER performance vs. XPD from 1 to 20 dB in the pedestrian and vehicular channels with $E_b/N_0$ = 16 dB. As can be seen, smaller XPD degrades performance. In this Figure we also plot the BER performance assuming perfect XPD knowledge at receiver (after cross polarization interference cancellation). Thus in order to enhance the performance in very low XPD conditions, methods to estimate and remove the cross-polarization interference could be investigated.

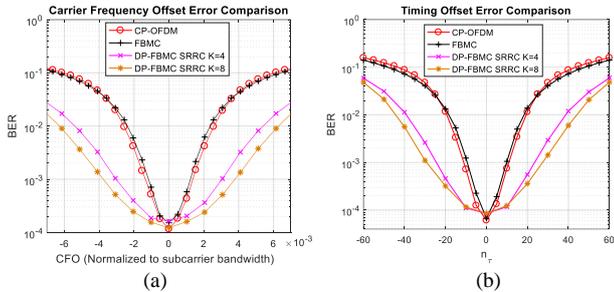

Figure 7. (a) BER vs. CFO, (b) BER vs. TO, AWGN channel, $E_b/N_0$=12 dB, 16-QAM, *M*=512.

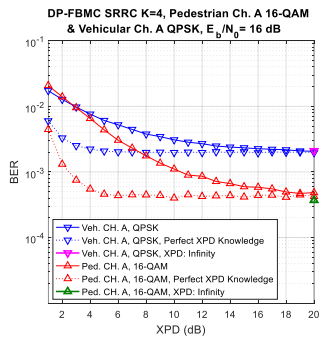

Figure 8. . BER vs. XPD, *DP-FBMC Structure I*, SRRC Filter with *K*=4, 16-QAM, ITU-R pedestrian and vehicular A channels, $E_b/N_0$ =16 dB.

## V. CONCLUSION

In this paper we proposed a novel FBMC system based on a dual polarization multiplexing technique. Using different time, frequency, and polarization multiplexing schemes and SRRC filters we showed that we can significantly suppress the FBMC intrinsic imaginary interference. In good XPD situations DP-FBMC shows significant advantages compared to conventional FBMC and CD-OFDM in multipath fading channels. DP-FBMC suffers when XPD is very small, therefore in future work XPD estimation and cancellation techniques could be investigated using pilot symbols.